# Using wearable proximity sensors to characterize social contact patterns in a village of rural Malawi

Laura Ozella[1], Daniela Paolotti[1], Guilherme Lichand[2], Jorge P. Rodríguez[1], Simon Haenni[2], John Phuka[3], Onicio B. Leal-Neto[2], Ciro Cattuto[1,4]

[1]ISI Foundation, Turin, Italy

[2]Department of Economics, University of Zurich, Switzerland

[3]College of Medicine, Lilongwe, Malawi

[4]University of Turin, Turin, Italy

## Abstract

Measuring close proximity interactions between individuals can provide key information on social contacts in human communities and related behaviours. This is even more essential in rural settings in low- and middle-income countries where the social well-being is not always guaranteed and there is therefore a need to understand contact patterns for the implementation of strategies for social protection interventions. Furthermore, understanding of social contact patterns that may shape the transmission of communicable diseases has utility in the design of control measures.

With the present study, we report the quantitative assessment of contact patterns in a village in rural Malawi, based on proximity sensors technology that allows for high-resolution measurements of social contacts. The system provided information on community structure of the village, on social relationships and social assortment between individuals, and on daily contacts activity within the village.

Our findings revealed that the social network presented communities that were highly correlated with household membership, thus confirming the importance of family ties within the village. Contacts within households occur mainly between adults and children, and adults and adolescents. This result suggests that the principal role of adults within the family, in particular adult women, is the care for the youngest. On the other hand, most of the inter-household interactions occurred among caregivers and among adolescents. We studied the tendency of participants to interact with individuals with whom they shared similar attributes (i.e., assortativity), in particular we tested the influence of age and gender. Age and gender assortativity were observed in inter-household network, showing that individuals not belonging to the same family group prefer to interact with people with whom they share similar age and gender. Age disassortativity is instead observed in intra-household networks. Family members congregate in the early morning, during lunch time and dinner time. In contrast, individuals not belonging to the same household displayed a growing contact activity from the morning, reaching a maximum in the afternoon.

The data collection infrastructure used in this study seems to be very effective to capture the dynamics of contacts by collecting high resolution temporal data and to give access to the level of information needed to understand the social context of the village.

## Introduction

Describing close proximity interactions allows to create contact networks representing frequency of social contacts in human communities. Contact network analysis can be used to better understand social interaction patterns and related behaviours (Borgatti et al., 2009; Chami et al., 2014) and the transmission of diseases (Funk, et al., 2010; Danon et al., 2011). Collecting high resolution data on the contact rates between individuals is a major challenge in most settings, particularly in rural low and middle income areas. Furthermore, many infectious diseases have emerged or re-emerged in rural low and middle income settings in the last century (Fenollar et al., 2018). Including observed contact data, of these harder to reach populations, in stochastic models of transmissible diseases could help better predict epidemics and has utility in the design of preventive and control measures such as vaccination and social distancing (Mossong et al., 2008; Salathé et al., 2010).

Contact diaries have usually been used to record contact information in low and middle income countries. Few studies have investigated social contact patterns in the Asian continent. Social networks in household-structured communities have been estimated in Vietnam (Horby et al., 2011) and in rural India (Johny et al., 2017) through the use of paper-based questionnaires. In Africa, a robust literature documents the use of contact diaries to understand and quantify households' connectivity (Hosegood and Timaeus, 2005; Cassidy and Barnes, 2012) and social contacts relevant for diseases transmission (Johnstone-Robertson et al. 2011; Crampin et al., 2008; Chami et al., 2014; Kiti et al. 2014). Specifically, in Malawi, Rock et al. (2016) have studied the social networks and social participation of youth living in extreme poverty in rural sites, and Helleringer and Kohler (2007) have investigated the relationship between the sexual networks of young adult population of several villages and the position of HIV-positive individuals within these networks. However, such survey-based approach permits obtaining only an approximate estimate of the number of contacts and of their durations, especially when repeated over many days or weeks. Moreover, young or illiterate participants may face difficulty in comprehending the process of completion of the questionnaire and need the assistance of trained workers (Johnstone-Robertson et al. 2011).

Recently, the use of proximity sensing technology to collect social contact information has emerged as an alternative. Mobile phones have been used to continuously collect proximity information within rural communities by Bluetooth devices within the scanning range of the user (typically 5–10 m) (Yoneki and Crowcroft, 2014). On the other hand, wireless proximity sensors can gather proximity interactions less than 1-2 meters separations distance every few seconds (Cattuto et al., 2010) in an objective and unsupervised way. These proximity events are relevant to detect social interactions, such as conversations, or spread of infections through physical touch or via aerosol. Proximity sensors were successfully deployed in rural Kenya to characterize contact patterns that may shape the transmission of respiratory infections in schools and households (Kiti et al., 2016; Kiti et al., 2019).

In this study, we report on the use of wearable proximity sensors to measure face-to-face proximity and pattern of contacts between individuals in a village in rural Malawi. The main objectives of the study were: (1) to assess the differences in contact patterns intra-household and inter-households, as well as their structural and temporal heterogeneities; (2) to study the social structure of the village at a group-level, by analysing the community structure, and at individual-level, by measuring the patterns of social assortment (specifically gender and age assortativity).

## Materials and Methods

### Data collection

The present study is part of a wider project on Child and Youth Development Study funded by UNICEF Malawi. The data collection was conducted between 16[th] December 2019 and 10[th] January 2020 in Mdoliro village in Dowa district in the Central Region of Malawi. The data were obtained and processed using a proximity-sensing application have been described in detail in several previous works (e.g., Cattuto et al., 2010; Stehlé et al., 2011; Ozella et al., 2018), and more recently it was used to detect proximity events between animals (Wilson-Aggarwal et al., 2019; Ozella et al., 2020). This system is based on wearable proximity sensors that exchange ultra–low power radio packets (Cattuto *et al*., 2010). Sensors in close proximity exchange with one another a maximum of about 1 power packet per second, and the exchange of low-power radio-packets is used as a proxy for the spatial proximity of the individuals wearing the sensors (Cattuto *et al*., 2010). In particular, close proximity is measured by the attenuation, defined as the difference between the received and transmitted power. In this study, we set the attenuation threshold at -75 dBm to detect proximity events between devices situated within 1-1.5 m of one another (e.g., Ozella et al. 2019). This distance between individuals allows detection of a close-contact situation during which social interactions might occur and a communicable disease infection might be directly transmitted. A 'contact event' between two individuals was identified when the devices exchanged at least one radio packet during a time interval of 20 sec. After a contact is established, it is considered ongoing as long as the devices continue to exchange at least one radio packet for every subsequent 20 s interval. Conversely, a contact was considered broken if a 20 sec interval elapses with no exchange of radio packets (Stehlé *et al.*, 2011; Kiti et al., 2019). Each device has a unique identification (ID) number that is used to link the information on the contacts established by the individual carrying the device. For the present study, the system was operated in a distributed fashion: contact data were stored in the local memory of individual devices. After collecting the devices at the end of the study, data from individual devices were downloaded, and the (temporal) contact networks recorded by individual devices were combined together to build a time-resolved proximity graph. In addition to contact information, each device periodically logs its orientation in space as measured by a tri-axial accelerometer.

The participants wore a sensor enclosed in a pouch and pinned to the front of a blouse/shirt in order to detect close-range proximity. The low-power radio frequency in use cannot propagate through the human body, and the position of sensor favours capturing face-to-face interactions. In addition, it was used the app Survey CTO through a tablet to collect metadata on the individuals included gender, age and which household they belonged to. A household was defined as the group of people living in the same house and eating from the same kitchen (Hosegood and Timaeus, 2005). Participants were grouped into three age categories: < 10 years old (children), 11-18 years old (adolescents), and > 18 years old (caregivers). Training sessions were conducted with Health Surveillance Assistants (HSAs) and volunteers in the use of sensors and how participants should have worn it over the study period, and HSAs visited the village in order to check if participants were wearing the sensors properly.

### Ethical aspects

Only participants who have given their written consent (documented) were included in the research. In the case of children, consent is obtained from their guardians. In case of adolescents, consent is obtained from both themselves and their guardians. The study was approved by Ethical Committee at the University of Zurich (OEC IRB #2018-046) and Ethical Committee at College of Medicine in Malawi (P.10/19/2825).

### Data analysis

The proximity data were extracted from devices and cleaned by identifying anomalies in the recorded data that might point to sensors that were tampered with or suffered hardware/battery issues resulting in data loss or low-quality data. Participants were asked to remove the sensor overnight. Night contacts were disregarded from the analysis by using the tri-axial accelerometer data to identify the time periods during which the sensor did not move. This also allowed us to identify the time periods during which the sensors were not worn by the participants. This data was also disregarded from the analyses.

### Network analysis and contact matrices

For each participant in the study, we computed the number of contact events and the duration of each contact. Time-aggregated, weighted contact networks were generated: nodes correspond to individuals, an edge between two individuals indicates that at least one contact event involving those individuals was recorded during the temporal aggregation window. The weight $w_{ij}$ of an edge between nodes $i$ and $j$ is defined as the cumulative duration of the contact events recorded between those individuals. Network edges are undirected and the weights on the edges are regarded as symmetric ($w_{ij} = w_{ji}$). The degree $k_i$ of a node $i$ in the above network corresponds to the number of distinct individuals with whom individual $i$ has been in contact. Contact matrices were generated based on daily number and on daily duration of contacts by age category.

### Daily activity profiles

We studied the daily activity profile of contacts among individuals, extracting the probability of observing a contact as a function of the time along the day. We computed these activity profiles for each household, splitted in intra-household and inter-household contacts. Additionally, we create two aggregated datasets that join all the timestamps of the observed contacts: the aggregated data for intra-household contacts and the aggregated data for inter-household contacts. We computed the Kolmogorov–Smirnov statistical distance $d_{KS}$ between the aggregated data of all the households and the data observed for each household (for intra-household daily activity profile), and of all individuals and the data observed for each individual (for inter-households daily activity profile). The Kolmogorov-Smirnov distance between two probability distributions *i* and *j* is defined as the maximum of the absolute difference between the cumulative density functions $C_i$ and $C_j$ of both distributions $d_{kS}(i,j) = max_\tau |C_i(\tau) - C_j(\tau)|$. This distance is bounded between 0, when the two compared distributions are identical, and 1, when the overlap between both distributions is null. In this particular case, we studied the distributions arising from the empirically observed contacts. The times of the contacts of each household (for intra-household data) and of each individual (for inter-households data) *i* are described by the cumulative density function $C_i$ given by $C_i(\tau) = \frac{1 - \Sigma_{t<\tau} N_i(t)}{\Sigma_t N_i(t)}$ where $N_i(t)$ is the number of observed contacts at time t, computed identically for the aggregated data. We considered a range of times of 24 h duration. We observe that the periodicity of this range may influence the measurement of dKS. For example, if we consider the origin of times at midnight and there are contacts between 23:00 and 01:00 in the aggregated dataset, but there is no contact at one specific household along this range, the maximum difference between the cumulatives would be influenced by the origin of times, such that this difference would be lower if the origin of times is included along the observed range. To tackle this issue, we generated the origin of times as a uniform random number between 0 and 24 h. We generated, for each pair of distributions, 100 different origins of times, computed the Kolmogorov-Smirnov distance for each origin of times, and the considered $d_{KS}$ was the minimum among all these samples.

**Community detection**

We used the Louvain algorithm (Blondel et al., 2008) to identify community structure in the aggregated networks. This method maximizes a modularity score for each community, where the modularity quantifies the quality of an assignment of nodes to communities. This means evaluating how much more densely connected the nodes within a community are, compared to how connected they would be in a random network. To see if community membership was determined by the participants' gender, age or household, we used the Normalized Mutual Information (NMI) score to scale the results between 0 (no mutual information) and 1 (perfect correlation).

**Assortativity**

We studied the assortativity (i.e., the tendency of the individuals to associate with individuals of similar characteristics) in gender and age-category. Our aim was to understand if individuals with same gender and same age-class will be more likely to interact, and if there are differences between intra and inter-households contacts. To test statistically the evidence of assortativity behaviour, we checked the measured assortativity with respect to a null model. We created an ensemble of realizations of a null model by shuffling nodes' attributes' and computing for those realizations the assortativity. Then, we computed the number of contacts and the total time in contacts (weights) between individuals in aggregated observed networks, and we compared the empirical results with the distribution of values obtained for the null model.

## Results

The whole study duration was 26 days (from 16$^{th}$ December 2019 to 10$^{th}$ January 2020), however, the proximity sensors were deployed and collected at different time within this time window. Therefore, each sensor had a different deployment period, and this varied from 16 days to 23 days (median 20 days). We included in the data analysis the contact data for all the participants when they wore the sensors simultaneously, and this overlapping deployment period was 13 days. On the other hand, we studied the intra-household contact data when the family members wore sensors simultaneously, and this overlapping deployment period was different according to the household (range 16 - 22 days).

**Network analysis and contact matrices**

**Intra-household contact data**

Overall, 28 households were included in the data analysis, accounting for a total of 84 sensors, distributed as follows: 17 children, 24 adolescents, and 43 caregivers. The mean number of household members wearing a sensor was 3 (range 2 – 5). We generated contact matrices based on duration of contacts (in seconds) and on number of contacts (contact events) by age class. We divided the total daily contact durations and the total daily number of contacts by the days during which the family members wore sensors simultaneously, and by the total number of persons belonging to an age class, thereby obtaining the daily mean of the contact durations and the daily mean of contact events *per capita* (Figure 1). The highest mean contact durations and mean number of contacts corresponded to the contacts between caregivers and children, and between caregivers and adolescents.

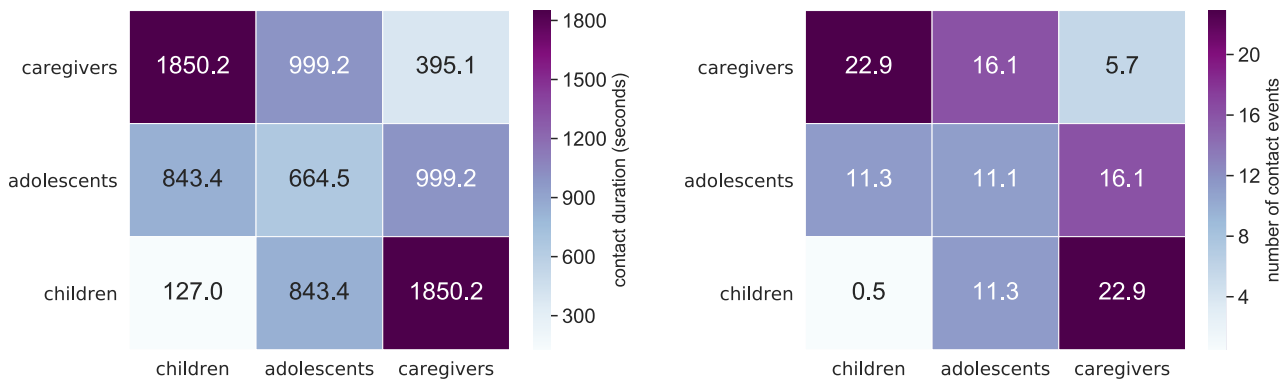

Figure 1. **Intra-household contact matrices**. Contact matrices giving the daily mean contact durations in seconds per capita (left panel) and the daily mean contact events per capita (right panel) by age class.

**Inter-household contact data**

The aggregated contact network for the participants that had contacts with individuals not belonging to their household had 74 nodes and 264 edges. The nodes are distributed as follows: 10 children, 24 adolescents, 40 caregivers (one of them had also the role of Health Surveillance assistant (HAS)), and one HSA was external from the village's population. All the adolescents involved in the study had contacts with individuals not belonging to their household. Overall, the median degree (i.e., number of connections with other individuals) was 6 (range 1 – 33). The median degree of children was 7 (range 1 -11) and of caregivers was 5 (range 1 - 33). The adolescents had the highest median degree: 8 (range 1 – 15). As we expected, the individual with highest number of connections was an HSA (degree 33), however the HSA external from the village had links with 10 people. We also generated contact matrices based on daily mean of the contact durations and daily mean of contact events *per capita* (Figure 2). The highest mean contact durations and mean number of contacts corresponded to the contacts among caregivers and among adolescents.

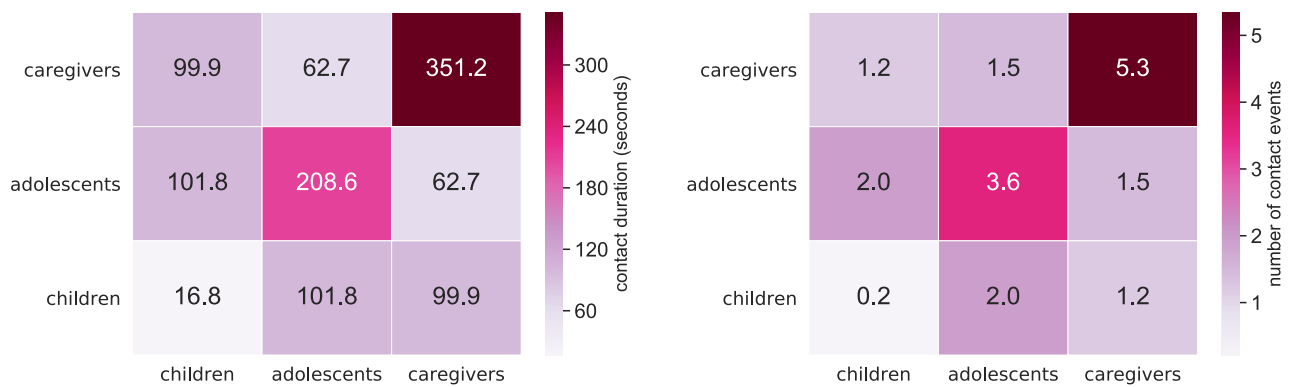

Figure 2. **Inter-household contact matrices**. Contact matrices giving the daily mean contact durations in seconds per capita (left panel) and the daily mean contact events per capita (right panel) by age class.

### Daily activity profiles

Intra-household contacts showed an intense growth during the early morning (from 5am) and two activity peaks around lunch time (1pm) and dinner time (7pm) (Figure 3). Likewise, inter-household contacts increased from 5am, but with a moderate intensity, and showed a peak of activity around 4pm (Figure 4).

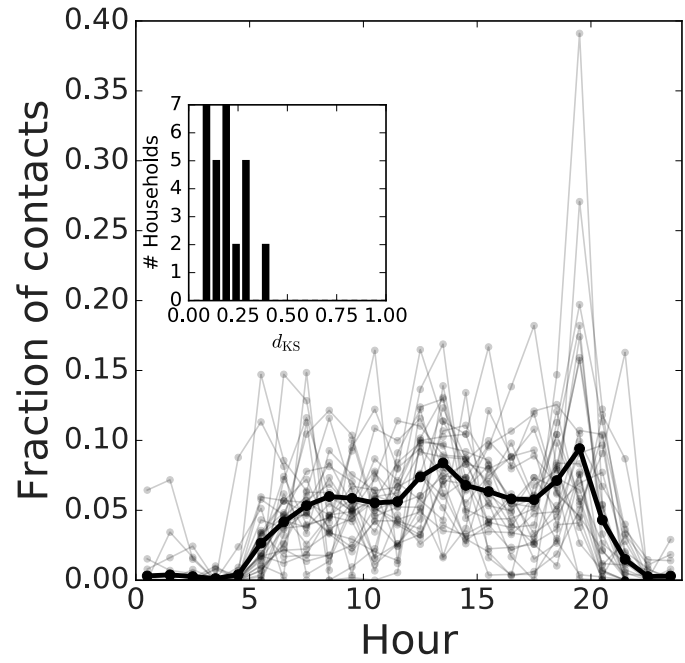

Figure 3. **Intra-household daily activity profiles**. Number of observed contacts happening at each hour of the day divided by total number of observed contacts. Each thin line corresponds to a different household and the thick line represents the aggregated dataset. Inset: number of households with a specific Kolmogorov-Smirnov distance $d_{KS}$ to the aggregated dataset.

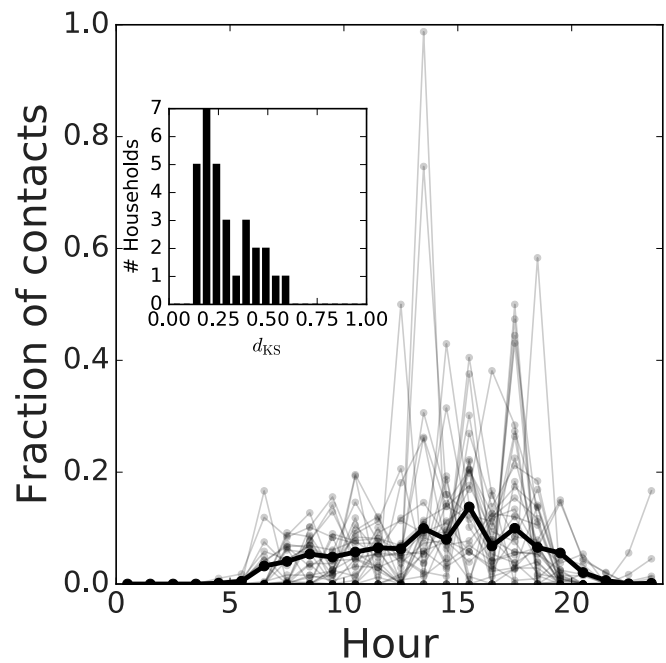

Figure 4. **Inter-household daily activity profiles**. Number of observed contacts happening at each hour of the day divided by total number of observed contacts. Each thin line corresponds to a different household and the thick line represents the aggregated dataset. Inset: number of households with a specific Kolmogorov-Smirnov distance $d_{KS}$ to the aggregated dataset.

### Community detection

We studied the community structure of the aggregated network by considering all the participants, both intra-household and inter-households contacts (86 nodes and 355 edges). Community analysis using Louvain's algorithm showed high modularity in the unweighted network (0.48) and the weighted network (0.75). Networks with high modularity have dense connections between the nodes within communities but sparse connections between nodes in different communities. The community membership was strongly correlated with household membership in both the unweighted network (NMI = 0.73) and weighted network (NMI = 0.82). Community membership had no significant relationship with either the individual's gender (unweighted NMI = 0.02; weighted NMI = 0.04) or age class (unweighted NMI = 0.03; weighted NMI = 0.11).

### 3.3 Assortativity

We measured the assortativity with respect to the gender and the age class of the participants. We considered the following hypothesis in relation to a null model: (*H0*) The observed fraction of contacts and fraction of time in contact involving individuals with same characteristic is compatible with a shuffled graph having exactly the same topological structure as observed empirically. We produced 1000 randomized equivalents of observed networks by shuffling individuals' characteristic. We computed the fraction of contact events and the fraction of total time in contact involving individuals with same gender and the same age class, intra and inter-households (Table 1).

Table 1. Fraction of contact events (unweighted) and the fraction of total time in contact (weighted) (on total number of contact events and on the total time spent in proximity, respectively) involving individuals with same gender and the same age class, intra and inter-households.

|  | gender | | age class | |
|---|---|---|---|---|
|  | unweighted | weighted | unweighted | weighted |
| *Intra-household* | 0.44 | 0.52 | 0.25 | 0.25 |
| *Inter-households* | 0.60 | 0.57 | 0.47 | 0.75 |

Then, we tested for the assortativity behaviour by computing the distribution of the fraction of contacts and fraction of total time in contact, linking nodes of same characteristics in the null model, and comparing it to the empirical value of the observed network of this ratio.

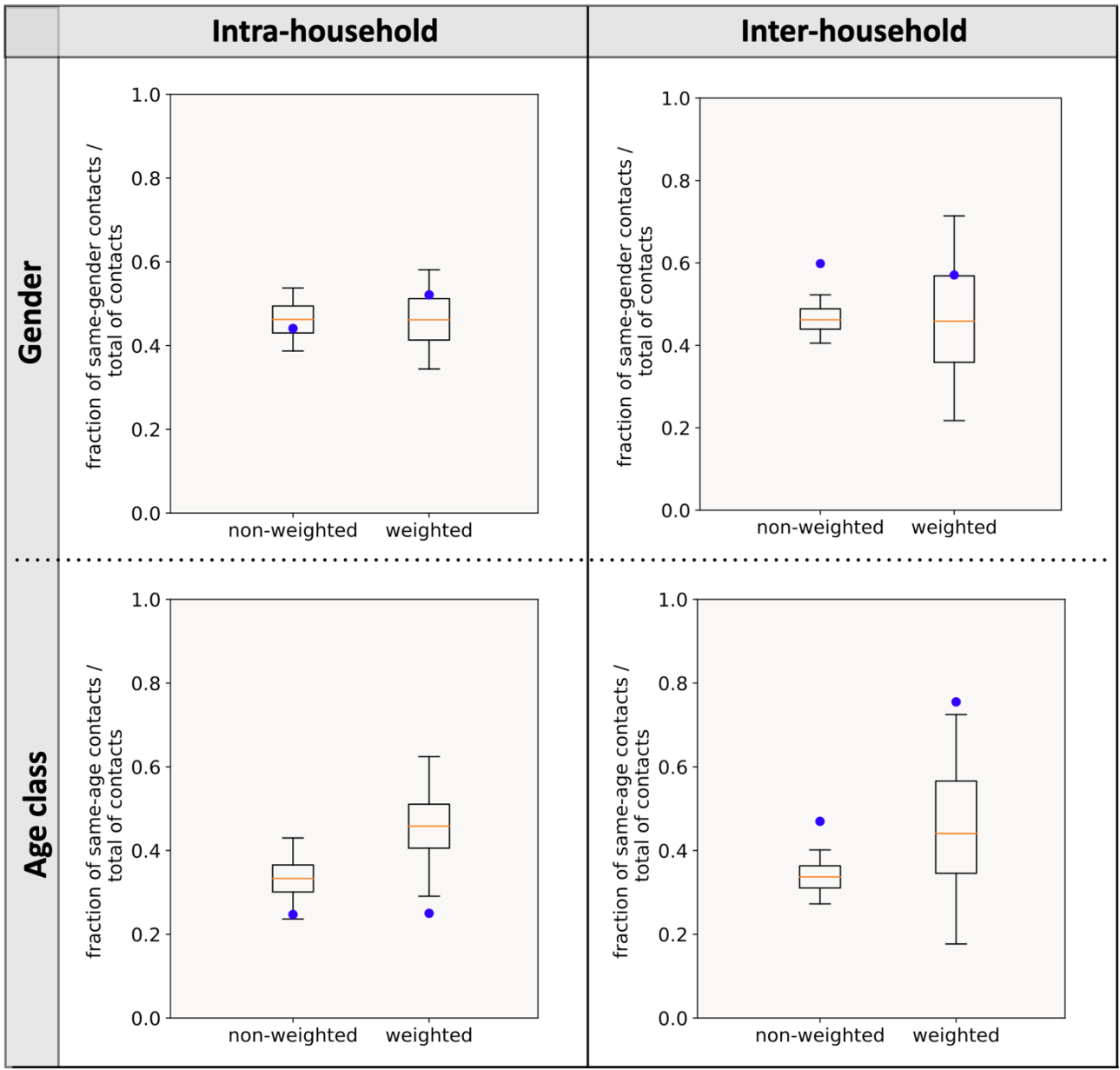

Figure 5. **Statistical test of gender and age assortativity for inter and intra household contacts** (non-weighted = fraction of contact events; weighted = fraction of total time in contact). Blue dots indicate the empirical values involving individuals of the same gender and age. Box plots show the distribution of the fraction of contact events (non-weighted) and fraction of total time in contact (weighted) resulting from the null model. In each box, the orange line marks the median and the extremities of the box correspond to the 25 and 75 percentiles: the whiskers give the 5 and 95 percentiles of each distribution.

Regarding the gender, we do not reject the null hypothesis H0 for unweighted (*p-value* = 0.680) and weighted contacts (*p-value* = 0.424) among individuals living in the same household, and for weighted contacts (*p-value* = 0.426) among individuals living in different households, the empirical values are included in the 95% confidence intervals of the null distribution. We can reject the H0 for unweighted inter-households contacts (*p-value* < 0.001), the empirical value is above the 95% confidence intervals of the null distribution, by demonstrating the preference of individuals living in different household to have contacts with individuals of same gender (i.e., gender assortativity) (Figure 5, upper panels).

Regarding the age classes, we do not reject the null hypothesis H0 for unweighted contacts (*p-value* = 0.06) among individuals living in the same households, the empirical values are included in the 95% confidence intervals of the null distribution. We can reject the H0 for weighted intra-household contacts (*p-value* = 0.006), the empirical value is below the 95% confidence intervals of the null distribution, by demonstrating the preference of individuals living in the same household to spend time with individuals of different age (i.e., age disortativity). We can also reject the H0 for unweighted and weighted inter-households contacts (p-values < 0.001 and 0.023, respectively), the empirical value is above the 95% confidence intervals, by demonstrating the preference of individuals living in different households to interact with individuals of the same age (i.e., age assortativity) (Figure 5, bottom panels).

## Discussion

With the present study, we report the quantitative assessment of contact patterns in a village in rural Malawi, based on proximity sensors technology that allows for high-resolution measurements of social contacts. The system provided information on community structure of the village, on social relationships and social assortment between individuals, and on daily contacts activity within the village, both intra and inter-households.

Our findings revealed that the social network presented communities that were highly correlated with household membership, thus confirming the importance of family ties within the village. The household is usually the fundamental social and economic unit in African villages, where individuals have more frequent and intense interactions (Hosegood and Timaeus, 2005), and where a significant part of children's development occurs (Bradley and Putnick, 2012). Families not only offer the access to basic necessities, such as food and shelter, but also safeguard a safe environment for young children and adolescents. The quality of care and parenting practices plays a key role in child growth (Engle et al. 2007) and a correct stimulation promote optimal child development through responsive and appropriate interactions with caregivers (Landry et al., 2006). Moreover, households play a key role in the transmission of infectious diseases, and household composition may influence transmission risks (Fraser, 2007; Horby et al., 2011). Nevertheless, a household is not an independent entity, but it is embedded in the broader structure of the village, where there are kinships and other relationships between individuals. Understanding and quantifying interactions both within and across households can offer a whole picture of social life. Children's early life experience is shaped not only by family contexts but also by the social ties formed within the village and social fabric characteristics play an important role in determining child development. We obtained contact matrices stratified by age-category on number of contact events and time spent in proximity. Our results showed a clear difference between intra and inter-household interaction patterns. Caregivers (adults more than 18 years old) had a greater number of contacts and time spent in proximity with children and adolescents living in the same household. This result suggests that the role of adults within the family is the care for the youngest. In addition to age, gender plays an important role in shaping the burden of youth care. In our study, adult women had more interaction with children and adolescents than male (see supplementary material). In developing countries, Female are more likely than male to take on caregiving activities, where mainly elderly women have to cope with the care of grandchildren and children (Schatz, 2007; Kalomo and Besthorn, 2018). However, we found that the role of male caregivers is not negligible, in particular in relation to children. Men in families represent an important resource for children's well-being. In West African countries, children are reared in large extended families with a clan-based kinship centered around a polygynous headman with little parental involvement (Nsamenang, 2010). Nonetheless, his role is crucial to guarantee a social position to the young since he provides social connections with the rest of the clan (Nsamenang, 2010).

On the other hand, our results showed that most of the inter-household interactions occurred among caregivers. Kiti et al. (2016) found similar results in rural Kenya, where most of the contacts and the total time spent in proximity across households are recorded between adults. We did not find relevant gender differences, and this shows that both adult men and adult women have social contacts outside the household. Even adolescents had high number of contact events and high time spent in proximity with individuals of the same age-category not belonging to their household. Moreover, adolescents had the highest median degree (i.e., number of distinct individuals with whom an individual has been in contact) in the inter-household network compared to the other age-categories, and this demonstrates the high sociality of youth individuals. Rock et al. (2016) studied how the sociality of adolescents can influence positively their mental and physical health in poor context in Malawi. However, a survey of sexual partnerships among young adults in several villages of Likoma, an island on Lake Malawi, showed that the high connectivity of social network leads to an increased risk of HIV infections (Helleringer and Kohler, 2007).

At an individual level, we studied the tendency of participants to interact with individuals with whom they shared similar attributes (i.e., assortativity), in particular we tested the influence of age and gender. Age and gender assortativity were observed in inter-household network, showing that individuals not belonging to the same family group prefer to interact with people with whom they share similar characteristics. Age disassortativity is instead observed in intra-household networks, this is easily explained by the parenthood and caregiving relationship between adults and youth within the family groups. Similarly, intra-household age assortativity was observed in rural Kenya by Kiti et al. (2016), in particular between children aged 6-14 years old, while age disassortativity was observed for adult groups.

We collected data continuously over 24 hours a day, and we obtained a clear trend of the temporal activity of the village. Family members congregate in the early morning, during lunch time and dinner time. These results agreed with those obtained in a study in rural Kenya (Kiti et al., 2016), thus suggesting a typical family behaviour in African rural villages. A growing contact activity from the morning reaching a maximum in the afternoon in individuals not belonging to the same household was observed. We suppose that people congregate in the afternoon for work or other engagements (e.g., play for the youth). However, we have contact information only for the study participants and we are not aware of any other potential contacts they had with people outside of the study.

The study has a number of limitations to be considered before findings can be generalized. Overall, it was performed based on data for a relatively small population and only for one village, and the high-resolution contact network data we used spans less than one month.  Despite these few limitations, the data collection infrastructure used in this study provided the level of information needed to understand the social context of the village in detail. Proximity sensors captured the dynamics of contacts by collecting high resolution temporal data characterized by timescales comparable with those intrinsic to social dynamics, without the influences of recall bias as instead is the case with paper diaries. Moreover, the proximity interactions detected by the sensors can provide key information on transmission opportunities of infections within the village. The successful transmission of an infectious disease in a population is dependent on many factors, and one of the key factors is the frequency of contacts among infected and susceptible individuals (Horby et al., 2011; Cauchemez et al., 2011). The collection of reliable proximity data can provide crucial information on network properties, such as the presence of super-spreaders who are more likely to spread infections based on the number and duration of their interactions. In particular, in rural areas of the developing world, access to healthcare is scarce, therefore, it is paramount to be able to understand transmission of communicable diseases and to identify and control measures such as vaccination and social distancing. For future work, it would be interesting to consider more inter-connected villages, in order to understand the social contacts and to simulate the spread of infection in a wider and more inter-connected community.

**Acknowledgements**

We want to thank the Center for Child Development team members: Nicoľo Tomaselli, Walter Dzerouniam, Jiajing Feng, Faith Milongo and the UNICEF Malawi Country Office and UNICEF Innovation.

## Supplementary material

**Comparison between genders**

We found that women (female caregivers) have stronger interactions with children and adolescents of the same households, compared to the male adults (Figure S1). While both female and male adults had more interactions with people of the same age category belonging to different families (Figure S2).

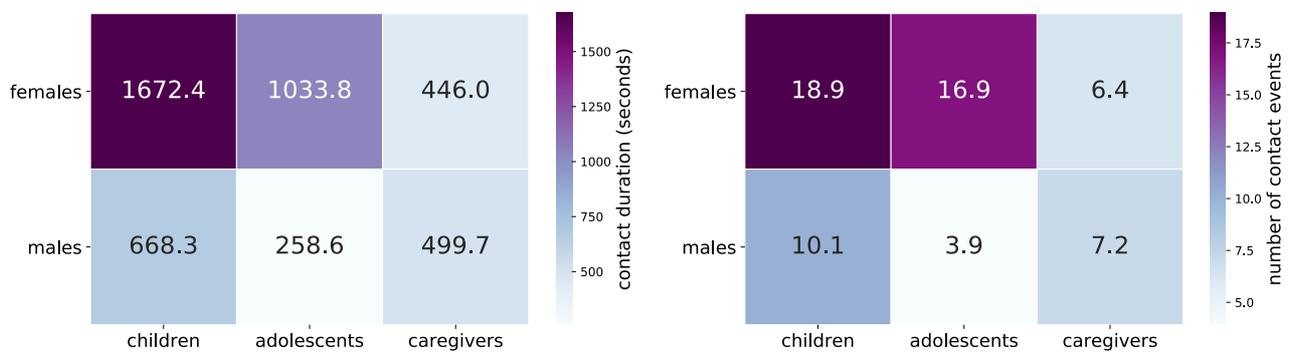

Figure S1. Intra-household contact matrices. Contact matrices giving the daily mean contact durations in seconds per capita (left panel) and the daily mean contact events per capita (right panel) of female and male caregivers.

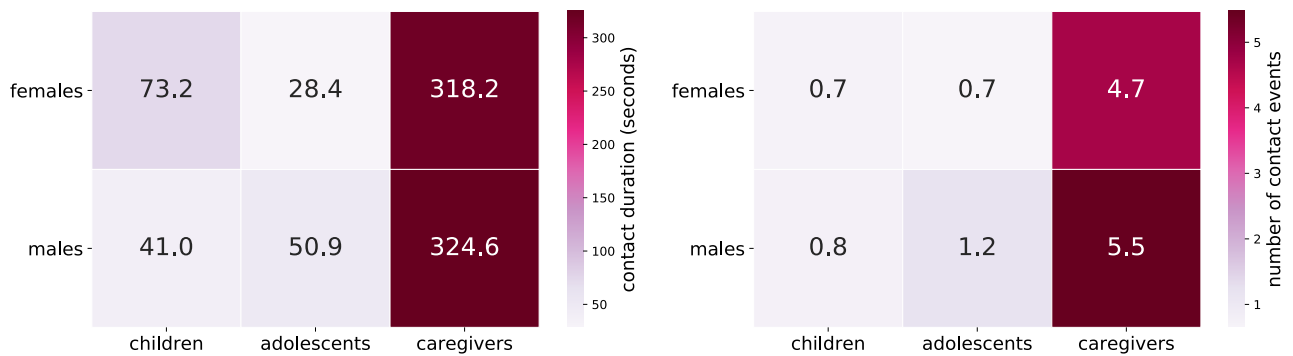

Figure S2. Inter-household contact matrices. Contact matrices giving the daily mean contact durations in seconds per capita (left panel) and the daily mean contact events per capita (right panel) of female and male caregivers.